# A finite element analysis model to predict and optimize the mechanical behaviour of bioprinted scaffolds

**Abhinaba Banerjee[1*], Sudipto Datta[2], Amit Roy Chowdhury[3], Pallab Datta[4]**

[1]Department of Mechanical Engineering, Indian Institute of Engineering Science and Technology, Shibpur, Howrah- 711103, West Bengal, India.

[2]Centre for Healthcare Science and Technology, Indian Institute of Engineering Science and Technology, Shibpur, Howrah- 711103, West Bengal, India.

[3]Department of Aerospace Engineering and Applied Mechanics, Indian Institute of Engineering Science and Technology, Shibpur, Howrah- 711103, West Bengal, India.

[4]Department of Pharmaceutics, National Institute of Pharmaceutical Education and Research, Kolkata 700054, West Bengal, India.

[*] Corresponding author: abhinaba1297@gmail.com

Bioprinting is an enabling biofabrication technique to create heterogeneous tissue constructs according to patient-specific geometries and compositions. Optimization of bioinks as per requirements for specific tissue applications is a critical exercise in ensuring clinical translation of the bioprinting technologies. Most notably, optimum hydrogel polymer concentrations are required to ensure adequate mechanical properties of bioprinted constructs without causing significant shear stresses on cells. However, experimental iterations are often tedious for optimizing the bioink properties. In this work, a finite element modelling approach has been undertaken to determine the effect of different bioink parameters like composition, concentration on the range of stresses being experienced by the cells in a bioprinting process. The stress distribution of the cells at different parts of the constructs has also been modelled. It is found that both bioink chemical compositions and stoichiometric concentrations can substantially alter the stress effects experienced by the cells. Similarly, concentrated regions of soft cells near the pore regions were found to increase stress concentrations by almost three times compared to the Von-Mises stress generated around the region of cells away from the pores. The study outlines the importance of finite element models in the rapid development of bioinks.

1. INTRODUCTION

Three-dimensional bioprinting has received considerable attention during the last decade due to its ability to produce patient-specific bioscaffold, biocompatible materials, and organ printing capabilities for transplantation. Researchers have been able to bioprint even complex vascularized tissue using multi-material 3D bioprinting [1]. Bioprinting of skin in vitro to improve skin function, treatment of skin disorder and considerable burn wound has been highlighted in the insightful review [2]. Other applications of bioprinting include fabrication of musculoskeletal tissues like bones and muscles [3], regenerative medicine [4], tendons and ligaments, cartilage tissue, etc. With the aid of computer-aided design (CAD) technique, computed tomography (CT) data, and using a photosensitive substance, three-dimensional (3D) printed artificial cornea has already been bioprinted in [5]. Attempts to contrive functional post-printing layered structures of brain neuronal tissue using peptide modified with enclosed cortical cell bioinks is a major accomplishment in bioprinting technology utilization [6].

Identification of proper bioinks that support tissue viability and proliferation required rheological and mechanical characteristics and biocompatibility is also a major challenge. Alginate sulphate combined with nanocellulose [7], agarose [3] [4], chitosan [10], collagen Type I [11], fibrin [7] [8], gelatin[14], hyaluronic acid [15], Matrigel$^{TM}$[16], methacrylate gelatin[17][18], poly(ethylene glycol) [19] are the most common types of natural and synthetic bioink materials used for scaffold-based bioprinting. Another emerging bioprinting technique is the scaffold-free printing using tissue spheroids directly as the building blocks [20]. But, both the methods are costly, time consuming, tedious and laboratory dependent [21][22]. Hence, numerical techniques and computer based simulations provide an alternate method to determine bioprintibility, biomimicry, structural and mechanical integrity of the bioink without even actually printing and performing repetitious extensive experiments.

Recently [23] developed a linear elastic finite element model and a non-linear regression model to study the effects of crosslinking volume and time on alginate scaffolds. Numerical methods to predict printability of nano fibrillar bioinks using rheology model, surface tension force model and quasi-dynamic contact angle model have been done in [24]. Determination of the tensile and flexural strength of bioprinted structures using finite element model has been recently done [25]. Researchers in [25] were able to successfully determine required material properties and select suitable layer thickness of bioprinted materials. Computational techniques are also used for failure criteria determination of biomaterials. Mechanical strength of 3D printed phantoms for abdominal aortic aneurysm has been assessed by finite element analysis in [26]. Computational modelling technique helped them [26] to precisely predict the aortic wall Von-Mises stress of geometrically complex "tissue-mimicking phantoms". Alginate printing from droplet optimization viewpoint and gelation mechanics has been highlighted using the reaction-diffusion model in the paper [27]. Bioinks are often subjected to high viscous forces during printing which affects the biomaterials been fabricated [28]. Computational models based on Navier's Stokes equations have been used for the determination of cell viability, droplet size, and mechanical impact during bioprinting in [29].

Almost all the modelling techniques developed to date have been done based on macro-scale analysis focussing only on the rheological properties of the bioink compositions, mostly due to limitations in computational resources and time complexity. Mechanical characterization of bioinks using microscale finite element modelling has not yet been reported in any research paper in the available literature. Although macroscale models require much lower computational time and are comparatively simpler to model, the simulation predictability of post bioprinting characteristics using microscale modelling is much accurate as they generate both precise and scalable structures. Intricate details like cell distribution characterization in hydrogels, multi-material structures, graded porosity distribution, and complex tissue structures in bioinks can also be taken into account by microscale finite element modelling with fidelity.

The purpose of the present study is to develop a microscale simulation and perform a comparative study of the mechanical response of alginate-HeLa cell bioprinted strand between experimentally observed data and finite element model. From the initial tensile test experiment of the strand structure, nonlinear hyperelastic material properties were formulated. Then using this material model a constitutive three-dimensional (3D) strand structure was modelled and analyzed in ANSYS software (Ansys Inc., V19 R2) implementing finite element technique. The response of the strand finite element model subjected to mechanical loading has been compared with the tensile test experiment data. A very good agreement is established between the two data during the validation of the microscale model. This technique of microscale simulation has been further extended to model four honeycomb structures made of GelatinMethacryloyl (GelMa) of concentration 10% and 15%, collagen and spider silk materials. All materials are having 5% (V/V) HeLa cell in them.

## 2. MATERIAL AND METHOD

### 2.1. Materials and alginate-HeLa cell solution preparation

Alginic acid sodium salt from brown algae (medium viscosity) was procured from Sigma-Aldrich at St. Louis, Missouri (USA). The specifications of the specimen used for the experiment were: Stock Keeping Unit (SKU): A2033, Pack size: 250G, Lot number: SLBQ3067V with Pcode: 1002311747. HeLa cell was purchased from National Centre for Cell Science(NCCS), Pune in India. The sample of alginate-HeLa cell as prepared, in which the alginate concentration was unaltered at 4% (W/V) in American Society for Testing and Materials (ASTM) Type III Grade water bought from Wasserlab, Navarra, Spain. The concentration of the HeLa cell for the strand specimens was 5% (V/V) of the alginate solution. HeLa cell was added, using a micro-pipette, to the alginate solution in a petri dish and stirred with a micro-tip to ensure uniform distribution of the cell in the alginate solution. The specimens that were used for the experiment are summarised in Table I. The alginate-HeLa cell solutions were freshly prepared and stirred for 24 hours in a mechanical stirrer prior to printing of the required structures by the bioprinter.

**Table 1. Shows the concentration of various components in the specimens tested.**

|  | Sodium Alginate salt (g) | ASTM Type III Grade water (ml) | HeLa cell per 5 ml of alginate solution (μl) |
|---|---|---|---|
| Specimen 1 | 4 | 100 | 0 |
| Specimen 2 | 4 | 100 | 250 |

### 2.2. Extrusion bioprinter

The 3D bioprinter used for printing of the alginate-HeLa cell composite is a custom-made 4-axis machine with X, Y, Z possible movements of the printer nozzle complemented by a syringe pump extruder. A stepper motor controls the X and Y axes movements which are driven on linear bearings whereas a lead screw drives the Z-axis movement. A geared stepper motor controls the syringe pump by stepping up the torque which is followed by an increase in extrusion pressure. To keep the Petri dish at a specified temperature a provision of a heated bed is also made. To achieve high precision while printing, the 32-bit architecture SAM controller is assisted with an automatic bed levelling and orthogonal compensation using the software. An open-source 3D printer slicing software application, CURA$^{TM}$, was used to slice the models followed by Pronterface software to issue G-Codes to guide our bioprinter.

### 2.3. Design of tensile test specimen

Cylindrical shaped strand of diameter 500μm and length 60mm was designed in Sketchup software (Trimble Inc., Sunnyvale, California) and saved in stereolithography (.stl) format. This was followed by the slicing of the 3D solid object in CURA$^{TM}$ version 15.04.4. The final structure as represented in **Fig. 1a**. was saved in G-Code format and then uploaded to the 3D printer for printing.

### 2.4. Extrusion printing of alginate-HeLa cell ink

The alginate-HeLa cell ink was then loaded into a syringe compatible with the printer. 3D printing was then performed with 50mm/sprint speed and a nozzle diameter of 0.5mm for the strand specimen. The temperature during the printing of the sample was fixed at 25°C. Alginate-HeLa cell ink was extruded on a glass petri dish followed by cross-linking of the structure using 1mL of 0.5 M $CaCl_2$ solution. The post-printing operation included washing the structures thrice with ethanol (95%, HimediaLaboratories, Mumbai, India) for 10 min followed by drying in a vacuum oven at a temperature of 40 °C.

### 2.5. Three-dimensional characterization of alginate-HeLa cell printed strand

The dimensions of the printed alginate-HeLa cell strand cannot be measured by any dial or digital Vernier callipers as they are not rigid bodies, rather viscoelastic in nature[30]. Therefore, a Nikon Digital camera (NIKON DS-QiMc, NIKON, Tokyo, Japan) attached to the microscope (Nikon ECLIPSE Ti) was used to acquire microscopic images of the strand. Strand length (SL) and perimeter (P)evaluation were done using "AxioVision SE64 Rel. 4.9.1" software (Carl Zeiss, Jena, Germany). Three images at 10x magnification were captured for each sample. Outer diameters obtained at different sections were 502.91 μm and 504.57 μm twice as shown in **Fig. 1b, 1c, and 1d**.

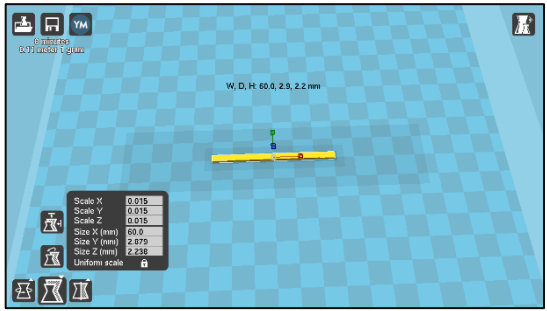
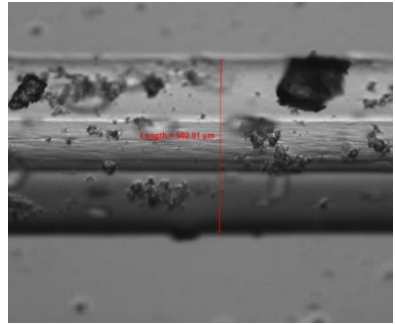

(a)                   (b)

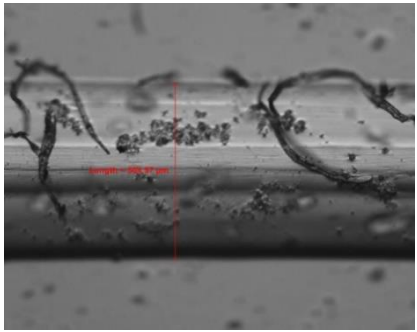
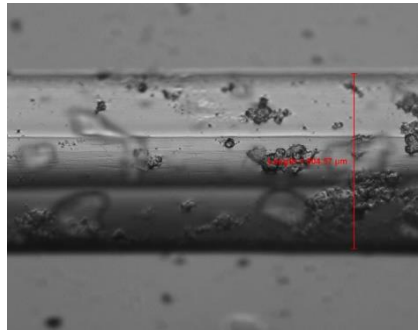
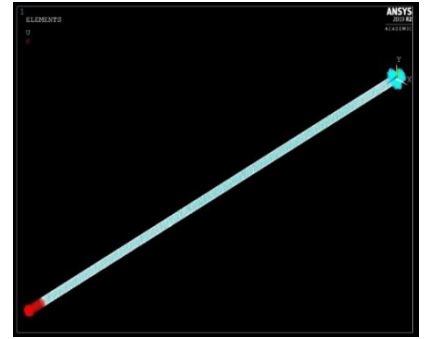

(c)             (d)             (e)

Figure 1. Diagram showing the single strand specimen:(a) Cura 3D model developed; (b), (c), (d) Images taken under the microscope; (e) Boundary conditions applied in finite element analysis

### 2.6. Mechanical characterization of alginate-HeLa cell specimens and data collection

A universal tensile testing machine (Tinius Olsen 5ST, Tinius Olsen, UK) was used to measure the tensile property of the samples. The cylindrical-shaped samples used for the test were of length 60 mm and diameter 500μm. The tests were performed at a load rate of 0.1 mm/min for Sample I and at 0.5mm/min for Sample II, with a load cell of 25 N for both the samples. The specimens were placed in a clip-type sample holder with 15 cm of the sample gripped inside the clamp on either side leaving 30mm as the gauge length. Each of the two alginate-HeLa cells and only alginate ink formulations were tensile tested. To check for the consistency of the strand each of the samples was subjected to ten preconditioning cycles of 10% strain prior to the sample testing. The specimens were elongated until failure occurred and the corresponding nominal stress was taken as the ultimate stress of the strand. All the tensile test data were collected using the "Horizon" software package (Tinius Olsen).

Before any analysis was done using the experimental data, it was smoothened in ORIGIN software (OriginLab Corporation, V2017) using the "locally weighted least squares" (LOWESS) technique due to the presence of a large number of data points[31]. The method first calculates weights of a center point and all surrounding points in the range using the "tri-cube weight function" given by:

$$w_i(x) = (1 - (\frac{|x - x_i|}{d_i})^3)^3$$

Here, $x$ is an adjacent point within the span of the current center point $x_i$ and $d_i$ is the distance along the abscissa from $x_i$ to the farthest points within the span. Next, the value at the current point is predicted by weighted linear regression. The same procedure is repeated for all the points of the experimental data.

## 2.7. Selection of constitutive material model for alginate-HeLa cell composite

Elastomeric materials that are subjected to large strain and nonlinear (elastic) deformation are modelled using strain energy density function based either on stress invariants or principal stresses.[32] The tensile test data of the alginate specimen without any HeLa cell was checked with the predefined hyperelastic material models available in ABAQUS software(Version 6.14) namelyNeo-Hookean; Ogden, N=1; Ogden, N=5; Reduced polynomial, N=2, and Arruda Boyce. The processed experimental test data were plotted with these material model curves and the best fit model was considered for further analysis. In this case Reduced polynomial, N=2 was selected to determine the alginate material constants as it was the closest approximation with test data.

The reduced polynomial form is a particular case of the polynomial form modelling of almost incompressible hyperelastic materials. The energy potential of the reduced polynomial model is given by:

$$U = \sum_{i=1}^{N} C_{i0}(\bar{I}_1 - 3)^i + \sum_{i=1}^{N} \frac{1}{D_i}(J_{el} - 1)^{2i} [33]$$

The material parameters for the HeLa cell were evaluated based on the parameters reported in the literature. The behaviour of the cell nucleus was assumed to be isotropic in nature with Young's modulus of 0.45MPa [34] and Poisson's ratio of 0.38, which lies well within the range of previously published data[35], [36].

## 2.8. Finite element modelling

The single strand was modelled in SolidWorks 2018 (Dassault Systèmes) as a cylindrical three-dimensional object with a diameter of 0.5mm and a length of 30mm. The Parasolid Model Part file (.X_T) of the model was imported in ANSYS and meshing was done. The entire volume was meshed using 10-node tetrahedral structural solid elements (SOLID 187) and a manual mesh size of 90μm was selected. The mesh size was restricted to such a small value as the average size of the HeLa cell is 30 μm. The 10-node tetrahedral element has the capability for simulating large deformations of fully incompressible hyperelastic materials and has a quadratic displacement behaviour. Next, the random distribution of the HeLa cell throughout the volume of the alginate hydrogel was incorporated by writing a script in Python. The element file and node file of the meshed solid was imported into the code after changes in material attributes were done the entire element attributes were saved in a new file. Next, the material constants obtained from ABAQUS software were included in ANSYS and the modified node file and element file were imported back. The boundary conditions were applied to the already meshed imported model as shown in **Fig. 1e**. One end surface of the cylindrical model was restricted of all degrees of freedom and on the other end surface, the tensile load was exerted. For analysis, the static large displacement option (NLGEOM key) was kept on as the material is hyperelastic and time stepping was applied.

## 2.9. Analysis of 3d bioprinted honeycomb structures and prediction of load distributions

After analysing the validity of the proposed technique of microscale modelling of bioink with experimental results, this modelling approach was extended to analyze four independent honeycomb structures made of GelMa (10% and 15%), collagen, and spider silk. Each of the materials also had a 5% (V/V) HeLa cell. The model is 0.625 mm thick with seven equally spaced regular hexagonal voids having an edge length of 1.25 mm. It was modelled in SolidWorks 2018 (Dassault Systèmes). The dimensions of the model were restricted to such small values due to computational capacity limitations, but the model can be easily replicated on a larger scale on having sufficient computational capacities.

The microscale pattern, like honeycomb or grooved surface, on bioprinted scaffold, is an important topological parameter for cell adhesion, proliferation, and cellular functional performance [37]–[41]. All these require the effective placing of the cells at specified positions on the scaffold. Two microscale models, one having coalesced cell near honeycomb cavity and the other away from the cavity, were also developed using the proposed microscale FE modelling technique, taking collagen as the scaffold material. The cell percentage in both cases was taken as 20% (V/V) with 18%(V/V) concentrated at specified places and the remaining 2% distributed throughout the rest of the volume. 6.81 N of force was applied on one end surface of the honeycomb structure in both cases. The coalesce cell position for both the models are shown in **Fig. 2a** and **Fig. 2b**. Microscale modelling of collagen honeycomb structure by varying cell volume percentage, from 5% to 10%, was also performed.

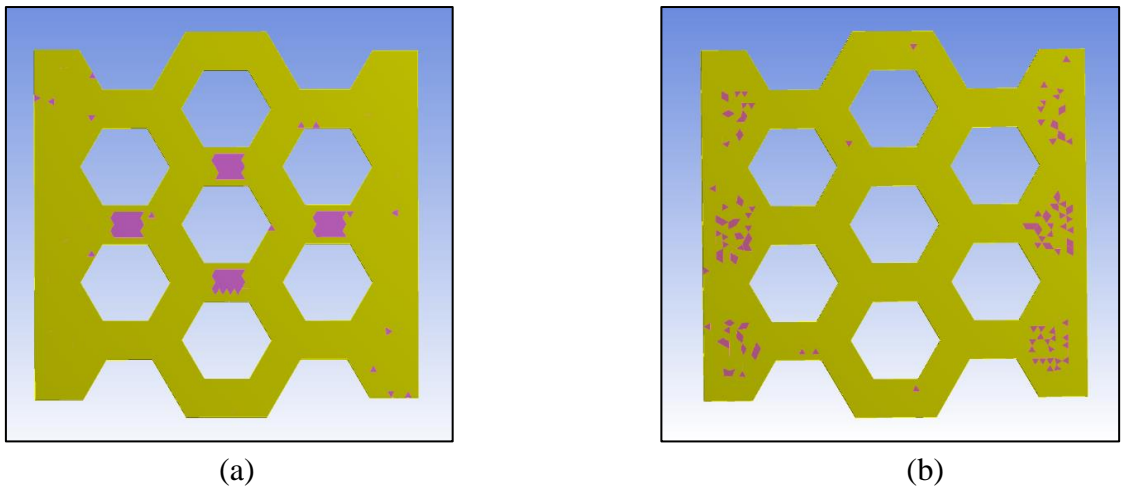

(a)            (b)

Figure 2. Illustration highlighting cell concentration in the finite element honeycomb model: (a) Near the cavity; (b) Away from the cavity

## 2.10 Selection of constitutive material model for honeycomb structures

The material properties of GelMa for 10% and 15% concentration were obtained from the literature [42]. The data points obtained from the stress-strain curve were further interpolated using WebPlotDigitizersoftware(Version 4.3). The expanded set of data points were used in ABAQUS software (Version 6.14) to select the best material model both for GelMa 10% and 15% as shown in **Fig. 3a** and **Fig. 3b**respectively. Ogden N=1 was found to be the best and most stable material model for both concentrations of GelMa.

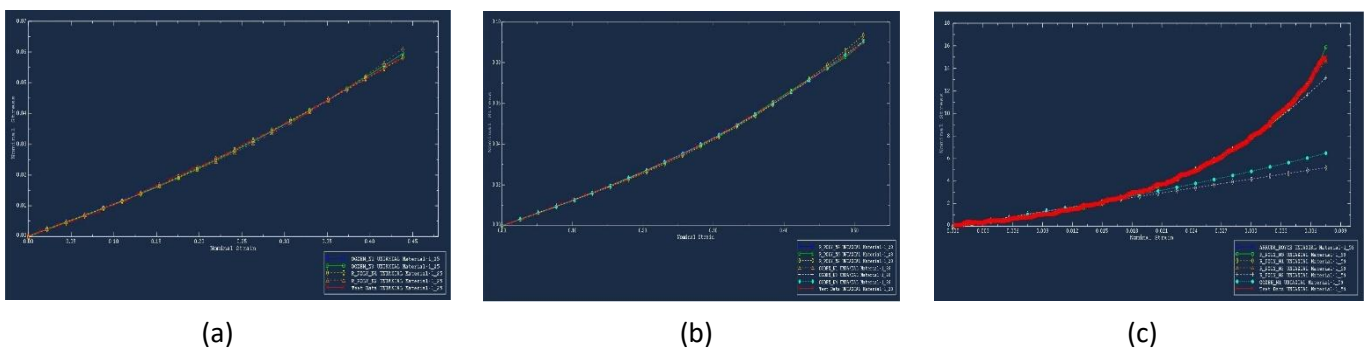

(a)            (b)            (c)

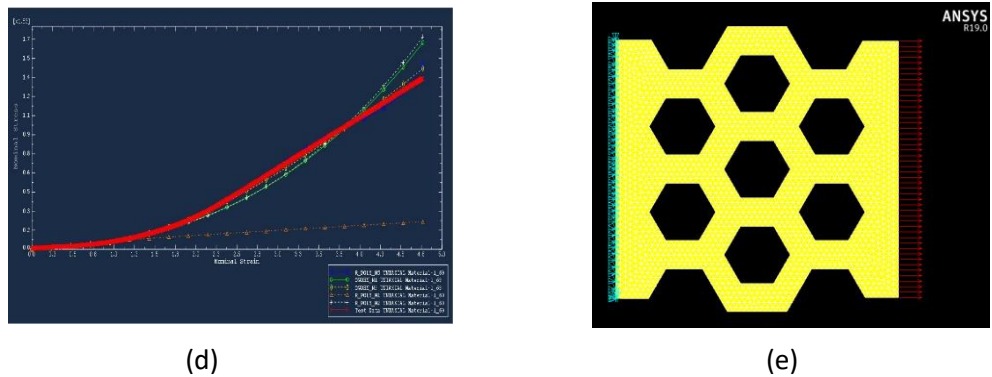

|         |         |
|---------|---------|
|   (d)   |   (e)   |

Figure 3. Diagram showing different material models for : (a) GelMa 10%; (b) GelMa 15%; (c) Collagen; (d) Spider Silk; (e) Boundary condition and load applied to honeycomb structure in finite element analysis.

The material property for collagen and spider silk was obtained from [43] and [44] respectively. The best material model for both was similarly obtained from ABAQUS software as in GelMa. Reduced Polynomial of order two was found to be the most stable material model for both materials, shown in **Fig. 3c** and **Fig. 3d**.

## 2.11  Finite element modelling of the honeycomb structure

The honeycomb model was imported to ANSYS in IGES formate for analysis. The entire volume was meshed using 10-node tetrahedral structural solid elements (SOLID 187) and a mesh size of 0.21 mm, assuming that 7 HeLa cells formed a cluster at a point. A total of 40,851 elements and 65,517 nodes were developed after meshing the model. The randomly distributed cell in the model was incorporated in a similar fashion as discussed earlier. The material property was defined as per the most stable material model for the respective materials as obtained from ABAQUS software. As the boundary condition, one end of the model was restricted all degrees of freedom and on the other end surface, a tensile load was exerted as shown in **Fig. 3e**. The large displacement option was selected for the analysis of all three models.

## 3.  RESULTS AND DISCUSSION

### 3.1.  Tensile test experiment

At first, the only alginate gel strand was tensile tested. The ultimate tensile force, just before the strand failed was found to be 0.145 N. The corresponding percentage directional strain along the strand length was found to be 16.3%. The directional stress as calculated by the "Horizon" software at failure was 0.722 MPa.

Next, the alginate HeLa cell strand specimen was tensile tested. The maximum tensile force that the strand could sustain just before failure was 0.175 N and the corresponding stress in the direction of strand length was 0.889 MPa. The ultimate percentage strain for this specimen was 21.57%.

### 3.2. Finite Element Analysis

Both tensile test experiments were replicated in finite element models one after another. The first finite element model, comprising of only Alginate material, could only be loaded up to a maximum tensile force of 0.142 N. The average Von-Mises stress in the model was calculated to be 0.725 MPa. The corresponding average directional strain(along strand length) in the model at the end of the analysis was 0.161 mm/mm.**Fig.4a** shows the stress-strain curve comparison of the experiment specimen and the finite element model for the only alginate gel material model.

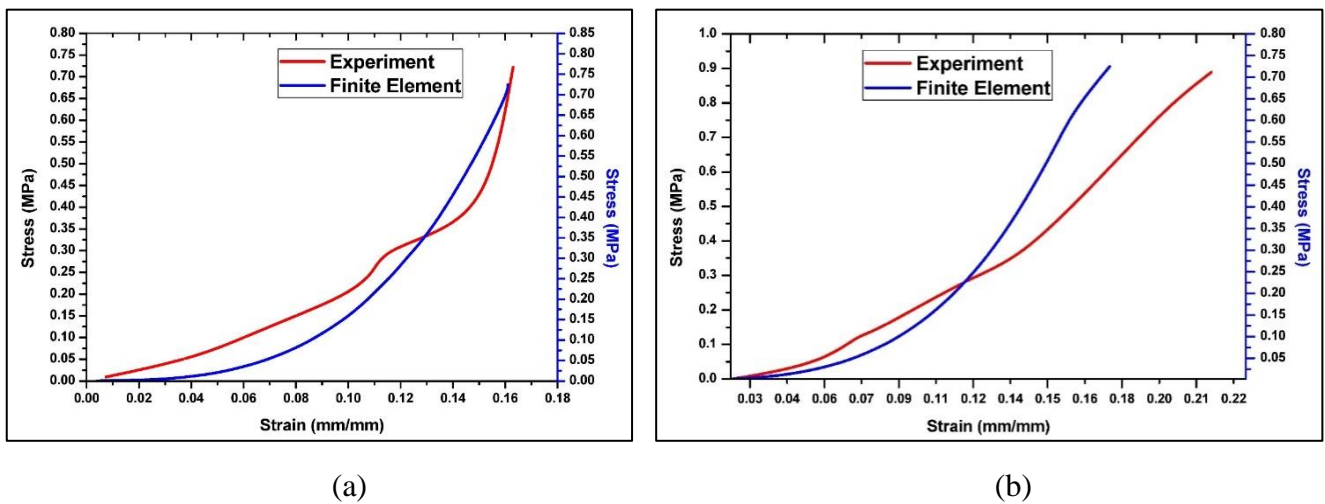

(a) (b)

Figure 4. Stress vs Strain curve for (a) Only alginate material specimen; (b) Alginate-HeLa cell composite material specimen.

The maximum tensile loading in the second model, a composite of alginate gel and HeLa cell, was limited to only 0.142 N. The corresponding average Von-Mises stress generated in the model was 0.725 MPa. The average strain along the applied load direction was obtained as 0.175 mm/mm. **Fig.4b** shows the stress-strain curve comparison of the experiment specimen and the finite element model for alginate gel and HeLa cell composite material model.

## 3.3. GelMa 10% - HeLa cell model analysis

The finite element model was loaded with a force of 0.25 N on one end. The deformed and undeformed shape at the end of the analysis is shown in **Fig. 5a**. The maximum Von-Mises stress generated in the model was 0.258 MPa. The corresponding Von-Mises strain developed was 45.2 %. The Von-Mises stress and strain distribution in the model is shown in **Fig. 5b** and **Fig. 5c** respectively. The stress versus strain curve for the model through the experiment is shown in **Fig. 5d**.

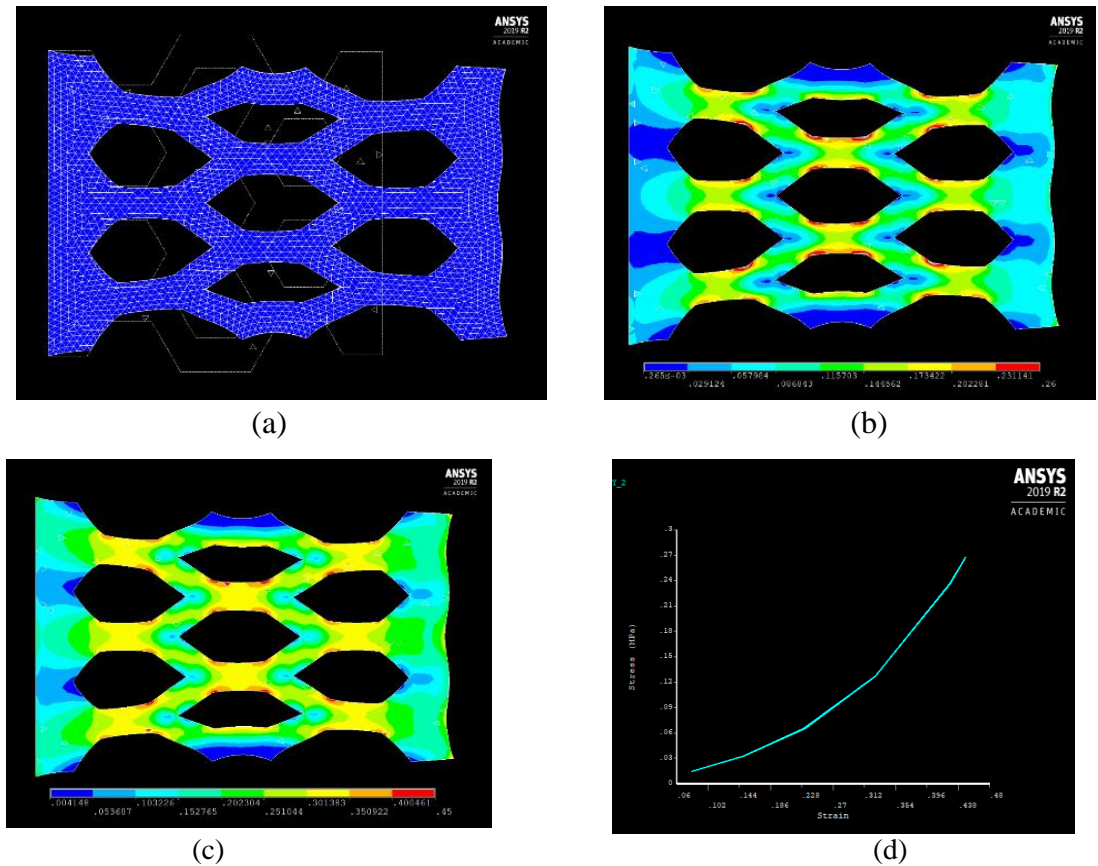

(a)          (b)

(c)          (d)

Figure 5. Illustrations related to GelMa 10% - HeLa cell model: (a) Deformed and undeformed edge after analysis;(b) Von-Mises Stress distribution;(c) Von-Mises Strain distribution; (d) Stress versus Strain curve.

## 3.2. GelMa 15% - HeLa cell model analysis

The finite element model was loaded with a force of 0.3023 N on one end. The deformed and undeformed shape at the end of the analysis is shown in **Fig. 6a**. The maximum Von-Mises stress generated in the model was 0.21 MPa. The corresponding Von-Mises strain developed was 52 %. The Von-Mises stress and strain distribution in the model is shown in **Fig. 6b** and **Fig. 6c** respectively. The stress versus strain curve for the model through the experiment is shown in **Fig. 6d.**

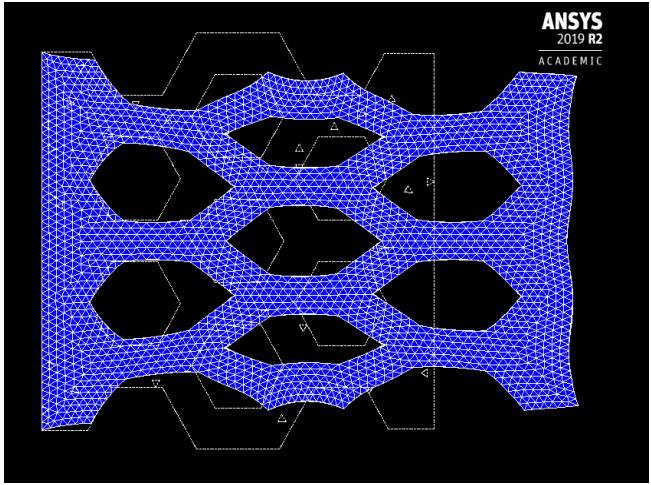
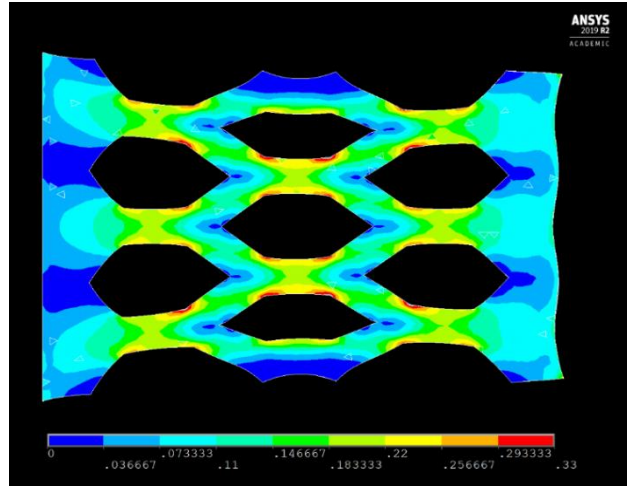

(a) (b)

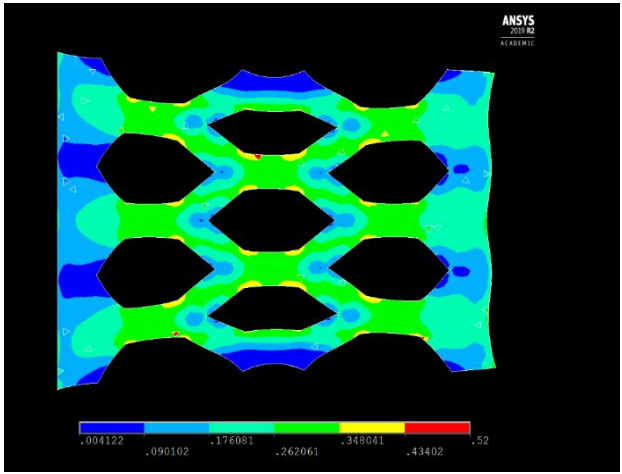
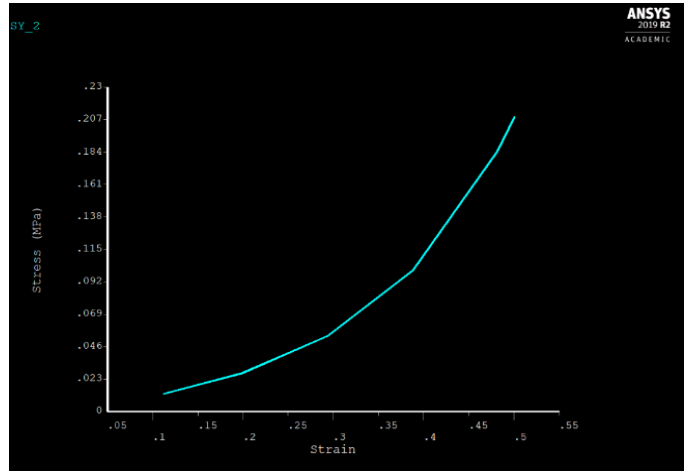

(c) (d)

Figure 6. Illustrations related to GelMa 15% - HeLa cell model: (a) Deformed and undeformed edge after analysis; (b) Von-Mises Stress distribution; (c) Von-Mises Strain distribution; (d) Stress versus Strain curve

### 3.3. Collagen - HeLa cell model analysis

The finite element model was loaded with a force of 70.53 N on one end. The deformed and undeformed shape at the end of the analysis is shown in **Fig. 7a**. The maximum Von-Mises stress generated in the model was 11.25 MPa. The corresponding Von-Mises strain developed was 3.82 %. The Von-Mises stress and strain distribution in the model is shown in **Fig. 7b** and **Fig. 7c** respectively. The stress versus strain curve for the model through the experiment is shown in **Fig. 7d**.

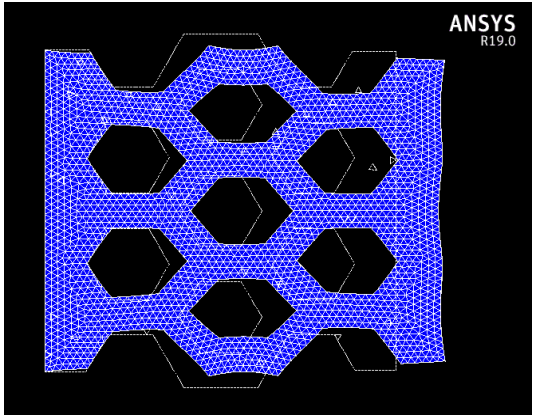

(a)

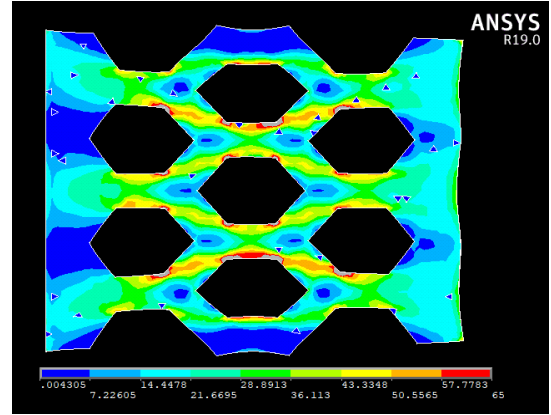

(b)

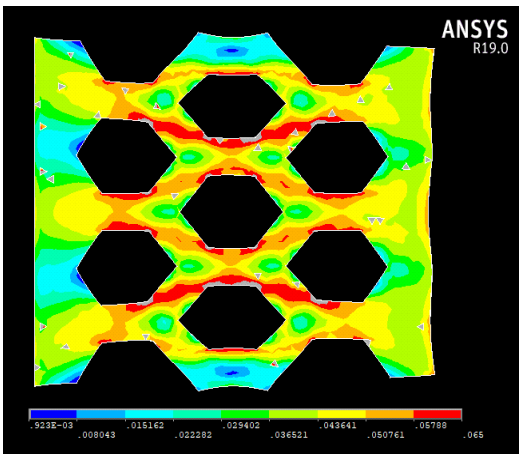

(c)

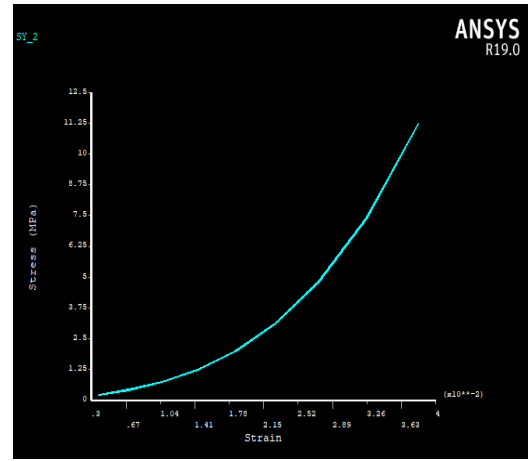

(d)

Figure 7. Illustrations related to Collagen - HeLa cell model: (a) Deformed and undeformed edge after analysis; (b) Von-Mises Stress distribution; (c) Von-Mises Strain distribution; (d) Stress versus Strain curve

### 3.4. Spider silk - HeLa cell model analysis

The finite element model was loaded with a force of 26.94 N on one end. The deformed and undeformed shape at the end of the analysis is shown in **Fig. 8a**. The maximum Von-Mises stress generated in the model was 24.054 MPa. The corresponding Von-Mises strain developed was 27 %. The Von-Mises stress and strain distribution in the model is shown in **Fig. 8b** and **Fig. 8c** respectively. The stress versus strain curve for the model through the experiment is shown in **Fig. 8d**.

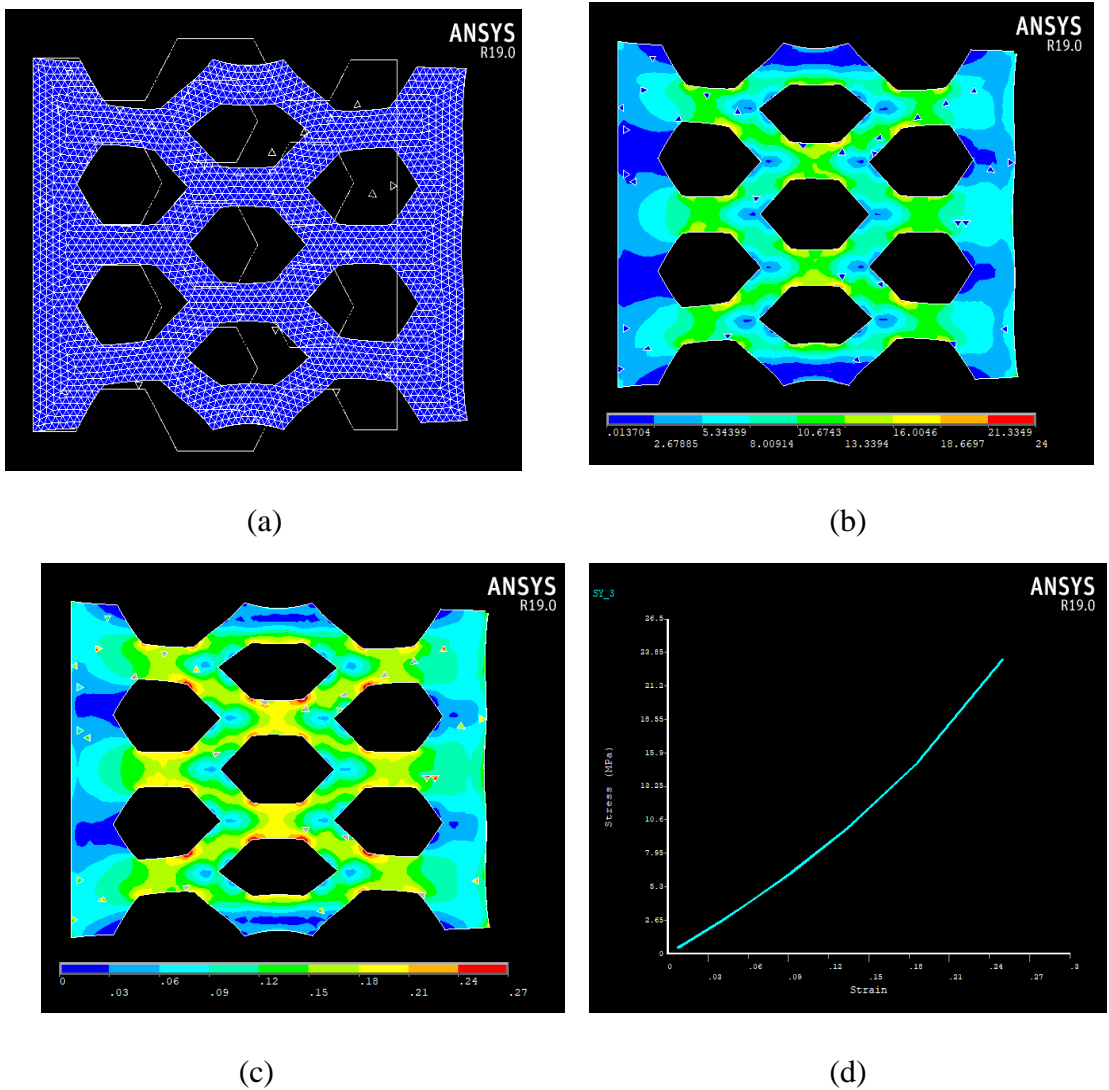

(a)                                     (b)

(c)                                     (d)

Figure 8. Illustrations related to Spider Silk - HeLa cell model: (a) Deformed and undeformed edge after analysis; (b) Von-Mises Stress distribution; (c) Von-Mises Strain distribution; (d) Stress versus Strain curve

The stress distribution in the honeycomb model shows a heterogeneous stress field for the region containing the cavities and an approximately homogenous stress distribution both at the fixed end (section a-a) and at the end where force is applied(section b-b), **Fig. 9a**. The maximum stress regions are observed at the regions highlighted in **Fig. 9b**. Section a-a (**Fig. 9a**) is also the region of minimum stress for the spider silk model. The maximum tensile stress for naturally spun silk is around 600 MPa, as obtained in experiments done by [45]. This ultimate stress limit reduces considerably, almost by a factor of 25 times, on adding 5% (V/V) HeLa cell with spider silk. Spider silk with cells cannot endure as large strain rates of 500% as possible for only spider silk polymer.

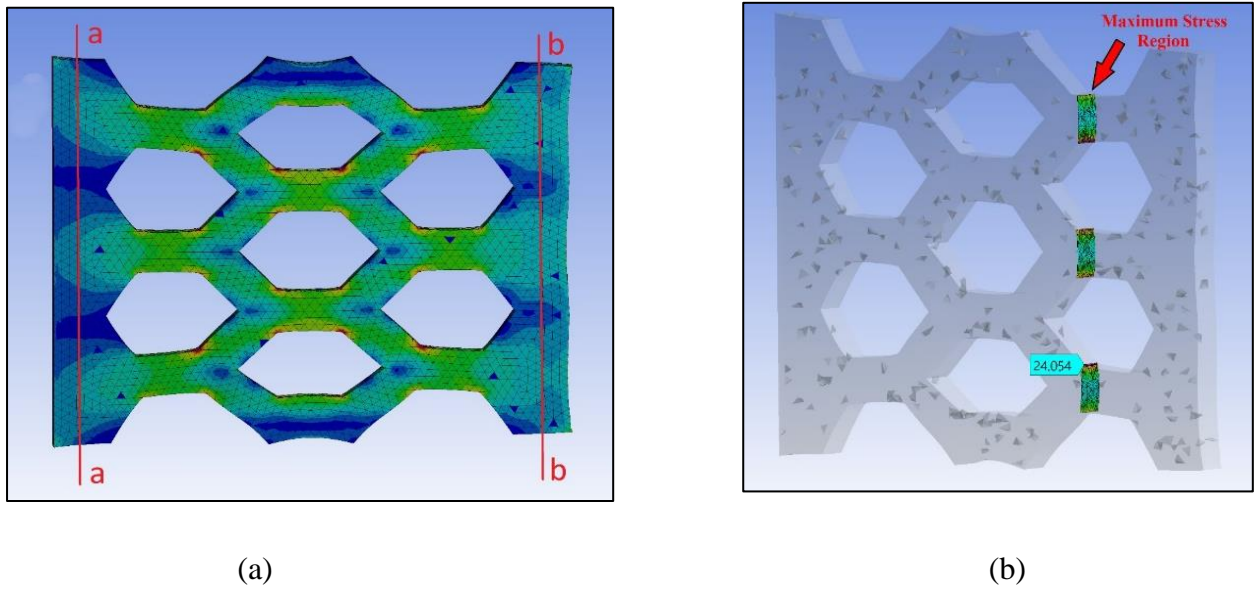

(a)                                                  (b)

Figure 9. Diagram showing various stress regions for spider silk model: (a) Sections of homogenous stress distribution; (b) Maximum stress locations.

Areas near the cavity are by default regions of high-stress concentration. The presence of concentrated regions of soft cells near the cavity increases the stress concentration further at those regions. When cells are present near the cavity the maximum Von-Mises stress experienced around that region is almost three times compared to the Von-Mises stress generated around the region of cells away from the cavity. The stress distribution at the periphery of the concentrated cell and the maximum stress generated for both the models is shown in **Fig. 10a** and **Fig. 10b**. This shows that an ideal distance of placing the cells from the cavity needs to be determined, while bioprinting, to ensure optimum stress generation depending on the specific application and placement position of the scaffold-cell structure.

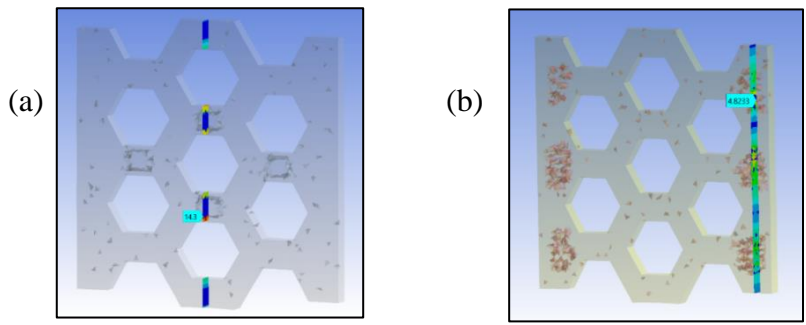

Figure 10. Maximum stress regions for coalesced cell present : (a) near cavity; (b) away from the cavity.

The cell percentage(V/V) was varied in the collagen-HeLa cell model between 5%, 10%, and 15%, and the associated changes in stress and strain were observed separately for regions close and far away from the honeycomb cavity. There was a marginal decrease of 2.27% and 1.5% in Von-Mises stress value for the increase in cell concentration from 5% to 10% and from 10% to 15% respectively, for regions at the cavity periphery. A similar decrease in stress value was observed for regions distant from the cavities and the corresponding decrease percentages in Von-Mises stress were 3.96% and 2.93% respectively. Although no significant changes in the strain value were observed for the model on changing the cell concentration both for regions close and far away from the cavity.

## 4. CONCLUSION

In conclusion, the study showed the development of realistic finite element models which can recapitulate the mechanical shear stresses experienced by the cells in a bioprinting process performed with different bioink compositions. The effect of different bioinks like Gel-MA, silk, collagen, and alginate used in different concentrations as well as different concentrations of the cellular component have been modelled. Moreover, the areas near the cavity are found to default regions of high-stress concentration. The presence of concentrated regions of soft cells near the cavity increased the stress concentration further at those regions as the maximum Von-Mises stress experienced around that region was almost three times compared to the Von-Mises stress generated around the region of cells away from the cavity. In bioink applications to print functional biological constructs, the cell concentration is generally kept at low percentages, as modelled here, which eventually proliferate with time. Hence the developed microscale model can be used for the rapid development of bioinks.

**REFERENCES :**

[1]     D. B. Kolesky, K. A. Homan, M. A. Skylar-Scott, and J. A. Lewis, "Three-dimensional bioprinting of thick vascularized tissues," *Proc. Natl. Acad. Sci.*, 2016.

[2]     C. Velasquillo, E. A. Galue, L. Rodriquez, C. Ibarra, and L. G. Ibarra-Ibarra, "Skin 3D Bioprinting. Applications in Cosmetology," *J. Cosmet. Dermatological Sci. Appl.*, 2013.

[3]     N. E. Fedorovich, J. R. De Wijn, A. J. Verbout, J. Alblas, and W. J. A. Dhert, "Three-Dimensional Fiber Deposition of Cell-Laden, Viable, Patterned Constructs for Bone Tissue Printing," *Tissue Eng. Part A*, 2008.

[4]     G. Gao and X. Cui, "Three-dimensional bioprinting in tissue engineering and regenerative medicine," *Biotechnology Letters*. 2016.

[5]     P. Xie *et al.*, "Application of 3-dimensional printing technology to construct an eye model for fundus viewing study," *PLoS One*, 2014.

[6]     R. Lozano *et al.*, "3D printing of layered brain-like structures using peptide modified gellan gum substrates," *Biomaterials*, 2015.

[7]     M. Müller, E. Öztürk, Ø. Arlov, P. Gatenholm, and M. Zenobi-Wong, "Alginate Sulfate–Nanocellulose Bioinks for Cartilage Bioprinting Applications," *Ann. Biomed. Eng.*, 2017.

[8]     L. Koch *et al.*, "Laser Printing of Skin Cells and Human Stem Cells," *Tissue Eng. Part C Methods*, 2010.

[9]     T. Xu, J. Jin, C. Gregory, J. J. Hickman, and T. Boland, "Inkjet printing of viable mammalian cells," *Biomaterials*, 2005.

[10]    Y. Yu, Y. Zhang, J. A. Martin, and I. T. Ozbolat, "Evaluation of Cell Viability and Functionality in Vessel-like Bioprintable Cell-Laden Tubular Channels," *J. Biomech. Eng.*, 2013.

[11]    A. M. Ferreira, P. Gentile, V. Chiono, and G. Ciardelli, "Collagen for bone tissue regeneration," *Acta Biomaterialia*. 2012.

[12]    M. Yanez, J. Rincon, A. Dones, C. De Maria, R. Gonzales, and T. Boland, "*In Vivo* Assessment of Printed Microvasculature in a Bilayer Skin Graft to Treat Full-Thickness

Wounds," *Tissue Eng. Part A*, 2015.

[13]     A. Skardal *et al.*, "Bioprinted Amniotic Fluid-Derived Stem Cells Accelerate Healing of Large Skin Wounds," *Stem Cells Transl. Med.*, 2012.

[14]     S. Sakai, K. Hirose, K. Taguchi, Y. Ogushi, and K. Kawakami, "An injectable, in situ enzymatically gellable, gelatin derivative for drug delivery and tissue engineering," *Biomaterials*, 2009.

[15]     Y. Luo, K. R. Kirker, and G. D. Prestwich, "Cross-linked hyaluronic acid hydrogel films: New biomaterials for drug delivery," *J. Control. Release*, 2000.

[16]     N. E. Fedorovich, H. M. Wijnberg, W. J. A. Dhert, and J. Alblas, "Distinct Tissue Formation by Heterogeneous Printing of Osteo- and Endothelial Progenitor Cells," *Tissue Eng. Part A*, 2011.

[17]     S. Wouter *et al.*, "Gelatin-Methacrylamide Hydrogels as Potential Biomaterials for Fabrication of Tissue-Engineered Cartilage Constructs," *Macromol. Biosci.*, 2013.

[18]     C. B Hutson *et al.*, "Synthesis and Characterization of Tunable Poly(Ethylene Glycol): Gelatin Methacrylate Composite Hydrogels," *Tissue Eng. Part A*, vol. 17, pp. 1713–1723, 2011.

[19]     M. Zhang, T. Desai, and M. Ferrari, "Proteins and cells on PEG immobilized silicon surfaces," *Biomaterials*, 1998.

[20]     V. Mironov, R. P. Visconti, V. Kasyanov, G. Forgacs, C. J. Drake, and R. R. Markwald, "Organ printing: Tissue spheroids as building blocks," *Biomaterials*, 2009.

[21]     J. Wieding, A. Wolf, and R. Bader, "Numerical optimization of open-porous bone scaffold structures to match the elastic properties of human cortical bone," *J. Mech. Behav. Biomed. Mater.*, 2014.

[22]     P. F. Egan, V. C. Gonella, M. Engensperger, S. J. Ferguson, and K. Shea, "Computationally designed lattices with tuned properties for tissue engineering using 3D printing," *PLoS One*, 2017.

[23]     S. Naghieh, M. R. Karamooz-Ravari, M. D. Sarker, E. Karki, and X. Chen, "Influence of crosslinking on the mechanical behavior of 3D printed alginate scaffolds: Experimental and numerical approaches," *J. Mech. Behav. Biomed. Mater.*, vol. 80, pp. 111–118, 2018.

[24]     J. Göhl, K. Markstedt, A. Mark, K. Håkansson, P. Gatenholm, and F. Edelvik, "Simulations of 3D bioprinting: Predicting bioprintability of nanofibrillar inks," *Biofabrication*, 2018.

[25]     D. Calneryte *et al.*, "Multi-scale finite element modeling of 3D printed structures subjected to mechanical loads," *Rapid Prototyp. J.*, 2018.

[26]     A. J. Cloonan, D. Shahmirzadi, R. X. Li, B. J. Doyle, E. E. Konofagou, and T. M. McGloughlin, "3D-Printed Tissue-Mimicking Phantoms for Medical Imaging and Computational Validation Applications," *3D Print. Addit. Manuf.*, 2014.

[27]     K. Pataky, T. Braschler, A. Negro, P. Renaud, M. P. Lutolf, and J. Brugger, "Microdrop printing of hydrogel bioinks into 3D tissue-like geometries," *Adv. Mater.*, 2012.

[28]     W. L. Ng, J. M. Lee, W. Y. Yeong, and M. Win Naing, "Microvalve-based bioprinting-process, bio-inks and applications," *Biomaterials Science*. 2017.

[29]     A. Tirella *et al.*, "Substrate stiffness influences high resolution printing of living cells with an ink-jet system," *J. Biosci. Bioeng.*, 2011.

[30]     D. D. Stromberg and C. A. Wiederhielm, "Viscoelastic Description of a Collagenous Tissue in Simple Elongation," *J. Appl. Physiol.*, 1969.

[31]     W. S. Cleveland and S. J. Devlin, "Locally weighted regression: An approach to regression analysis by local fitting," *J. Am. Stat. Assoc.*, 1988.

[32]     M. C. Boyce and E. M. Arruda, "Constitutive Models of Rubber Elasticity: A Review," *Rubber Chem. Technol.*, 2000.

[33]     L.-R. Wang and Z.-H. Lu, "Modeling Method of Constitutive Law of Rubber Hyperelasticity Based on Finite Element Simulations," *Rubber Chem. Technol.*, 2003.

[34]     S. Leporatti *et al.*, "Cytomechanical and topological investigation of MCF-7 cells by scanningforce microscopy," *Nanotechnology*, 2009.

[35]     D. Shin and K. Athanasiou, "Cytoindentation for obtaining cell biomechanical properties," *J. Orthop. Res.*, 1999.

[36]     F. Guilak, J. R. Tedrow, and R. Burgkart, "Viscoelastic properties of the cell nucleus," *Biochem. Biophys. Res. Commun.*, 2000.

[37]     H. Maleki *et al.*, "Mechanically Strong Silica-Silk Fibroin Bioaerogel: A Hybrid

Scaffold with Ordered Honeycomb Micromorphology and Multiscale Porosity for Bone Regeneration," *ACS Appl. Mater. Interfaces*, vol. 11, no. 19, pp. 17256–17269, 2019.

[38]	T. Kawano, M. Sato, H. Yabu, and M. Shimomura, "Honeycomb-shaped surface topography induces differentiation of human mesenchymal stem cells (hMSCs): Uniform porous polymer scaffolds prepared by the breath figure technique," *Biomater. Sci.*, vol. 2, no. 1, pp. 52–56, 2014.

[39]	H. Itoh, Y. Aso, M. Furuse, Y. Noishiki, and T. Miyata, "A honeycomb collagen carrier for cell culture as a tissue engineering scaffold," *Artif. Organs*, vol. 25, no. 3, pp. 213–217, 2001.

[40]	J. George, J. Onodera, and T. Miyata, "Biodegradable honeycomb collagen scaffold for dermal tissue engineering," *J. Biomed. Mater. Res. - Part A*, vol. 87, no. 4, pp. 1103–1111, 2008.

[41]	C. L. Gilchrist, D. S. Ruch, D. Little, and F. Guilak, "Micro-scale and meso-scale architectural cues cooperate and compete to direct aligned tissue formation," *Biomaterials*, vol. 35, no. 38, pp. 10015–10024, 2014.

[42]	M. Y. Shie, J. J. Lee, C. C. Ho, S. Y. Yen, H. Y. Ng, and Y. W. Chen, "Effects of gelatin methacrylate bio-ink concentration on mechano-physical properties and human dermal fibroblast behavior," *Polymers (Basel).*, vol. 12, no. 9, pp. 1–18, 2020.

[43]	J. A. J. Van Der Rijt, K. O. Van Der Werf, M. L. Bennink, P. J. Dijkstra, and J. Feijen, "Micromechanical testing of individual collagen fibrils," *Macromol. Biosci.*, vol. 6, no. 9, pp. 697–702, 2006.

[44]	T. Köhler and F. Vollrath, "Thread biomechanics in the two orb-weaving spiders Araneus diadematus (Araneae, Araneidae) and Uloborus walckenaerius (Araneae, Uloboridae)," *J. Exp. Zool.*, vol. 271, no. 1, pp. 1–17, 1995.

[45]	G. V. Guinea, M. Elices, J. Pérez-Rigueiro, and G. R. Plaza, "Stretching of supercontracted fibers: A link between spinning and the variability of spider silk," *J. Exp. Biol.*, vol. 208, no. 1, pp. 25–30, 2005.
21